%% file: ms.tex
\documentclass[10pt,conference,letterpaper,compsocconf]{IEEEtran}
\usepackage{epsfig}

\pdfoutput=1

\usepackage{xcolor}
\usepackage{url}
\usepackage{wasysym}
\usepackage{microtype}
\usepackage{flushend}

\usepackage{xspace}
\newcommand{\Prop}[1]{\underline{#1}\xspace}

\begin{document}

\title{Benchmarking Crimes:\\An Emerging Threat in Systems Security}

\author{\IEEEauthorblockN{Erik van der Kouwe\IEEEauthorrefmark{1},
Dennis Andriesse\IEEEauthorrefmark{2},
Herbert Bos\IEEEauthorrefmark{2},
Cristiano Giuffrida\IEEEauthorrefmark{2} and
Gernot Heiser\IEEEauthorrefmark{3}}
\IEEEauthorblockA{\IEEEauthorrefmark{1}\textit{Leiden University, The Netherlands}\\
\textit{e.van.der.kouwe@liacs.leidenuniv.nl}}
\IEEEauthorblockA{\IEEEauthorrefmark{2}\textit{Vrije Universiteit Amsterdam, The Netherlands}\\
\textit{$\lbrace$d.a.andriesse,h.j.bos,c.giuffrida$\rbrace$@vu.nl}}
\IEEEauthorblockA{\IEEEauthorrefmark{3}\textit{Data61 (formerly NICTA) and UNSW}\\
\textit{gernot@unsw.edu.au}}
}

\maketitle

\begin{abstract}
\input{sections/abstract.tex}
\end{abstract}

\input{sections/intro.tex}

\input{sections/benchcrimes.tex}
\input{sections/methodology.tex}

\input{sections/survey.tex}
\input{sections/impact.tex}
\input{sections/recommendations.tex}
\input{sections/limitations.tex}
\input{sections/relatedwork.tex}

\input{sections/conclusion.tex}

\appendix

\input{sections/surveyappendix.tex}

\footnotesize \bibliographystyle{IEEEtran}
\bibliography{bibliography,papers}

\end{document}

%% file: sections/abstract.tex
Properly benchmarking a system is a difficult and intricate task.
Unfortunately, even a seemingly innocuous benchmarking mistake can compromise
the guarantees provided by a given systems security defense and also
put its reproducibility and comparability at risk. This threat is
particularly insidious as it is generally not a result of malice and
can easily go undetected by both authors and reviewers.
Moreover, as modern defenses often trade off security for performance
in an attempt to find an ideal design point in the performance-security space,
the damage caused by benchmarking mistakes is increasingly worrisome.

To analyze the magnitude of the phenomenon, we identify a
set of 22 ``benchmarking crimes'' that threaten the validity
of systems security evaluations and
perform a survey of 50 defense papers
published in top venues.
To ensure the validity of our results, we perform the complete survey twice, with two independent readers.
We find only a very small number of disagreements between readers, showing that our assessment of benchmarking crimes is highly reproducible.

We show that benchmarking crimes are widespread even in papers
published at tier-1 venues.
We find that tier-1 papers commit an average of
five benchmarking crimes and we find only a single paper in our sample
that committed no benchmarking crimes.
Moreover, we find that the scale of the problem is constant over time,
suggesting that the community is not yet addressing it despite the problem
being now more relevant than ever.
This threatens the scientific process, which relies on reproducibility
and comparability to ensure that published research advances the state of the art.
We hope to raise awareness of these issues and
provide recommendations to improve benchmarking quality and safeguard
the scientific process in our community.

%% file: sections/intro.tex
\section{Introduction}
Benchmarking is essential in systems security---to compare different solutions
and reproduce prior results. At every program committee meeting for every
top venue in our field, heated discussions revolve around the question
whether the performance numbers reported in papers X and Y are reliable
and how they relate to each other. Making the wrong call is bad,
as nobody wants to accept or reject papers for the wrong reasons.
And after we accept a paper, we want to be able to reproduce and
compare the results in a meaningful way.
Faults in benchmarking are popularly referred to as
\emph{benchmarking crimes}~\cite{gernot},
even if authors generally do not commit them
intentionally\footnote{\emph{Crime} here has no ``criminal"
connotations, but is intended as a hyperbolic term following the original terminology
introduced by Heiser~\cite{gernot}.}.\looseness=-1

Bluntly speaking, benchmarking crimes threaten the validity
of the research results in publications. The obvious question then is:
how safe are we as a community from this threat?
And if we are not safe, how serious is this threat, and how can we mitigate it?
Phrased differently, we want to know how well the systems security
research community detects anomalies in benchmarking
in evaluation sections of papers published in tier-1 venues,
what the consequences are of false negatives,
and how to fix these ``vulnerabilities''.

In the community, there is wide agreement that performance benchmarks
are important to advance the field~\cite{sim2003using}.
In systems, it is clear that all security mechanisms incur
some performance overhead~\cite{wagner2015high}.
The aim is to keep the overhead as low as possible,
while raising the bar for attackers as high as possible.
Given an unlimited performance budget, techniques to build secure systems
under common threat models are already well-established---memory safety
being a typical example~\cite{softbound,cets}.
As a result, modern systems security research focuses on practical defenses (such as control-flow integrity~\cite{cfi}
or randomization~\cite{readactor}), that trade off
some security to achieve realistic perfomance guarantees.

Given these constraints, performance benchmarking is increasingly important
in systems security. Proper benchmarks allow us to compare different solutions
and reproduce research results. Improper benchmarking,
on the other hand, may set unrealistic standards and hamper progress in the
area.

In this paper, we take a closer look at benchmarking crimes
in systems security. While it would be good to also benchmark the
security of a solution, doing so in an unbiased way is much
harder~\cite{bellovin2006brittleness} and this paper primarily focuses on
performance benchmarking of defenses
(expanding on other dimensions when appropriate).
After discussing the objectives of performance benchmarking in general,
we carefully explore all the pitfalls that authors may encounter
when assessing the performance of their research artefact.
For each of these benchmarking crimes, we explain the negative impact
they may have on the validity or usefulness of the evaluation.

Finally, we assess the state of benchmarking in systems security.
We selected systems defense papers from USENIX Security,
the leading conference in systems security,
but also from the other tier-1 computer security venues
where systems security defenses are routinely published
(Security \& Privacy, CCS, and NDSS).
We finecombed some 50 papers and analyzed how they fared with respect to
the benchmarking crimes. For this purpose, we selected \emph{all} defense papers
with benchmarking results published in 2010 and 2015.
As nearly all papers in our data set have at least some benchmarking issue
(and many have several) and we found no clear difference between
the more recent and the older papers, we conclude that improper benchmarking
is a serious threat with little improvement in recent
years. Moreover, our analysis shows that more and more papers are affected
over time, confirming the increasing relevance of benchmarking crimes in our
community.

It is explicitly not our intention to point fingers.
As mentioned, all of the papers that we investigated exhibited some flaws
and we freely admit that some of our own past papers are no exception.
The point that we want to make is that the problem is not
with individual papers, but with the field.  
While we acknowledge that following all the guidelines is difficult
and sometimes impossible under time pressure,
we found that for many common and serious crimes the extra effort
is very low and our goal is to formulate specific guidelines moving forward
for the systems security community.
Moreover, by having concrete guidelines for benchmarking,
it is possible to build automatic tools that set up and run benchmarks
in such a way as to avoid benchmarking crimes.
We believe this is a promising area for future work.

We also refrain from speculating about the cause of the crimes we
encountered. Informal discussions at PC meetings frequently blame
the pressure to publish in good venues that lead authors to cut corners.
The reasoning is as follows. All defense solutions in systems security
represent a tradeoff between security and performance.
As a result, researchers frantically try to minimize the performance overhead,
while not compromising the security---sometimes doing whatever it takes
to stay under (fairly arbitrary) thresholds.
For example: ``The instrumentation overhead should be
under 5\%''~\cite{sok-memory-errors}. We do not deny such pressure exists,
but we have found no evidence of deliberate cheating.
We believe that most, if not all, benchmarking crimes we found are
unintentional and just denote insufficient attention devoted
to performance benchmarking in our community. As mentioned earlier, many
prevalent benchmarking crimes we found can, in fact, be prevented altogether
with little effort and simple benchmarking practices.
Our goal is to raise awareness of this increasingly important issue
and foster high-quality benchmarking to improve reproducibility
and comparability of research in our community.

\subsection*{Contributions}

This paper makes the following contributions:
\begin{itemize}
\item We raise awareness of a number of common pitfalls that affect
      the validity of benchmarking results in systems security.
      We report on 22 benchmarking crimes which are commonly found
      in systems research papers.
\item We present a survey of defense research papers in top security venues
      in recent years to demonstrate the impact of these benchmarking crimes
      in the systems security community.
\item We propose best practices to reduce the impact
      of improper benchmarking practices and improve
      the scientific process in our community.
\end{itemize}

%% file: sections/benchcrimes.tex
\section{Benchmarking Crimes}
\label{sec:crimes}

Almost every paper in computer systems requires an evaluation
that determines whether and how well the presented system achieves its goals.
One important purpose of the evaluation is to compare against other work:
it should show that the system improves the state of the art in some way and
allow possible later papers to show that they improve this system.
To allow for comparison, an evaluation must meet a number of requirements.
First of all, it should be \Prop{complete} in the sense that it verifies all
claimed contributions of the system and shows the extent of any negative impact
the system may have. All the presented results must be \Prop{relevant}
in the sense that they actually tell the reader something meaningful about
the system. Another important characteristic is
\Prop{soundness}, the requirement that all numbers measure what is intended
with reasonable accuracy and repeatability.
Finally, a general principle of science, requires papers to be
\Prop{reproducible}. That is, the information provided in the paper should
be sufficient to allow others to build the system and perform its evaluation
in the same way as the original.
A good paper should meet all these requirements, but unfortunately experience
shows that this is often hard to come by in practice.
Indeed, we found that most papers commit a number of \emph{benchmarking crimes}
that violate these properties.

In this section, we describe the benchmarking crimes we identified
and explain their importance. Our list is based in large part
on a reference web page by Heiser~\cite{gernot},
but we alter and add a number of benchmarking crimes and also
perform a systematic and large-scale survey of systems defense papers
at top conferences (see Section~\ref{sec:survey}) to determine whether
these benchmarking crimes are common in published papers in systems security.
We will show that the applicability of these crimes is not limited to
the operating systems community, but also extends to other subfields
of computer systems, in particular systems security.
This is particularly important because,
as we shall see, Heiser's original web page~\cite{gernot} published
in 2010 had insufficient impact in the systems security community.
Benchmarking crimes are still widespread and their relevance has,
in fact, grown over time.

We placed the 22 benchmarking crimes we identified in groups and
assigned codes (a letter for the group plus a number for the specific crime)
to simplify later references to them. We summarize the identified
benchmarking crimes and their impact in Table~\ref{tab:impact}.
While many crimes  impact multiple requirements, we merely show the \emph{most} affected ones. We describe the groups and the individual benchmarking crimes
in the following subsections and later elaborate on their impact
in Section~\ref{sec:survey}.

\begin{table*}[t]
  \caption{Benchmarking crimes and their impact; $\CIRCLE$=high-impact crime, $\ocircle$=other crime\label{tab:impact} (indicating only the \emph{most} affected requirements).}
\centering
\resizebox{6.5in}{!}{
\input{tables/impact.tex}
}
\end{table*}

\subsection{Selective benchmarking}

There is no single number that can fully express how well a system performs.
Performance overhead is multidimensional as different operations are affected in
different ways. For example, a system that performs CFI~\cite{cfi} instruments
indirect branches but leaves other operations alone. Therefore, it is likely to
incur substantial overhead for programs and workloads that perform many function
calls, especially if they are indirect (e.g., common C++ programs), but it will
incur minimal overhead if the program spends most of its time in a loop that
calls no functions. This has several implications for benchmarking, and when a
paper does not consider these implications it might result in a performance
evaluation becoming anywhere from slightly inaccurate to completely meaningless.

The first implication is that we should always make sure we include benchmarks
that perform all the kinds of work where one might reasonably expect an impact
on performance. If a system improves one kind of workload compared to the state
of the art but slows down another, it is important to show this to uncover
tradeoffs and allow readers to decide whether this solution is actually faster
overall.
If a paper does not include such benchmarks, it commits benchmarking crime
\emph{A1: not evaluating potential performance degradation}.
A typical example would be a system that instruments some system calls in the
kernel (potentially slowing them down) but runs only workloads that primarily
perform user mode computations.
In this case, the benchmarking results would be meaningless and would not allow
the reader to determine whether the system is practical or how it compares to
related work. This crime results in a lack of \Prop{completeness}.

Another implication is that, whenever a paper summarizes performance
as a single number, it must take care to ensure this number is representative
of real-world workloads. A number of benchmarking suites, such as
SPEC CPU2006~\cite{henning2006spec}, have been created for this purpose.
Different subbenchmarks stress different
types of operations and therefore result in different overhead numbers.
Any paper which arbitrarily selects a subset of benchmarks and
presents it as a single overall performance overhead number as if it
is still representative commits benchmarking crime
\emph{A2: benchmark subsetting without proper justification}.
If the missing subbenchmarks happen to be those that incur most overhead,
the overall performance number will be meaningless because important
components are missing (lack of \Prop{completeness}) and misleads the reader
into thinking the system performs better than it actually does (lack of
\Prop{relevance}). A typical example would be a system that instruments memory
management operations (potentially slowing them down) and omits the
memory-intensive \emph{perlbench} from SPEC CPU2006~\cite{henning2006spec}.
This problem is not limited to performance benchmarks;
a subset arbitrarily selected from a large set of tests is unlikely
to be representative of the full set regardless of whether they
benchmark performance or, for example, vulnerabilities that
the system attempts to mitigate.

Finally, benchmark configurations are often flexible and allow performance
to be measured in different settings. A typical example would
be the number of concurrent connections for a server program.
Since this configuration parameter is likely to affect overhead,
it is important to measure a range of concurrency settings.
Papers that fail to test performance over an appropriate range of settings
are guilty of benchmarking crime
\emph{A3: selective data set hiding deficiencies}.
For example, if throughput seems to scale linearly with the number of
concurrent connections, it suggests that the range of this variable is
too restricted because the system cannot keep this up forever.
Like the other two crimes in this group, it has the potential to result
in numbers that do not accurately reflect the practical performance impact
of the system (lack of \Prop{completeness}).

\subsection{Improper handling of benchmark results}

Our second group of benchmarking crimes deals with the question whether papers
interpret benchmarking results in appropriate ways.
Even when running the right benchmarks, the presentation of their results
can be misleading if they are processed in incorrect ways.
This group contains five benchmarking crimes related to incorrect handling
of benchmark results.

Microbenchmarks measure the performance of a specific aspect of the system.
While such benchmarks have value to determine whether a system succeeds
to speed up these particular operations, as well as for drilling down
on performance issues, they are not an indication of how
fast the system would run in practice. For this purpose, more realistic
macrobenchmarks are needed. Misrepresenting the results of microbenchmarks
is classified as
\emph{B1: microbenchmarks representing overall performance}
and threatens \Prop{relevance} because the presented results are misleading.

Benchmarks usually run either a fixed workload to measure
its runtime or repeat operations for a fixed amount of time to measure
throughput. One common mistake is for papers to consider the increase
in runtime or decrease in throughput to be the overhead.
However, for many workloads the CPU is idle some of the time,
for example waiting I/O. If the CPU is working while it would
otherwise have been waiting, this masks some of the overhead because
it reduces the CPU time potentially available for other jobs. 
A typical example would be a lightly loaded server program (e.g., at 10\% CPU)
that reports no throughput degradation when heavily instrumented,
given that the spare CPU cycles can be spent on running
instrumentation code (at the expense of extra CPU load). Papers that ignore this
effect are guilty of \emph{B2: throughput degraded by x\% \(\Rightarrow\) overhead is x\%}.
One possible way to avoid this crime is to ensure the CPU is fully loaded
by running a sufficient number of concurrent jobs. Alternatively, the
change in CPU load must be taken into account, e.g.\ by quoting the
cost of processing a certain amount of data.
When a paper commits this crime, it threatens the \Prop{soundness} of the results
and almost certainly results in the presented
overhead being lower than the actual overhead.

Another crime in this group is \emph{B3: creative overhead accounting}.
We use this to refer to any kind of incorrect computations
with overhead numbers. Examples include the use of percentage points
to present a difference in overhead, such as the case where the difference
between 10\% overhead and 20\% overhead is presented as 10\% more overhead,
while it is actually 100\% more (i.e., 2x). Another example would be computing
slowdown incorrectly, for example presenting a runtime that changes
from $5s$ to $20s$ as a 75\% slowdown ($1-\frac{5}{20}$)
rather than a 300\% slowdown ($\frac{20}{5} - 1$).
In all such cases, this crime results in presenting numbers
that are incorrect and therefore \Prop{unsound}.

When measuring runtimes or throughput numbers, there is always random variation
due to measurement error. If these measurement errors are large,
it typically means that there is a problem with the experimental setup
and the numbers measured should be taken with a grain of salt.
For this reason, we consider the lack of some indication of variance,
such as a standard deviation or significance test to be benchmarking crime
\emph{B4: No indication of significance of data}.
We classify this as a lack of \Prop{completeness} because without knowing
the amount of variation one cannot tell what the measured results really mean.

Papers that use benchmarking suites generally present a single overall
overhead figure representing average overhead.
Some authors use the arithmetic mean to summarize such numbers.
However, this is inappropriate because the arithmetic mean over
a number of ratios depends on which setup is chosen
as a baseline~\cite{fleming1986not} and is therefore not a reliable metric.
Only the geometric mean is appropriate to average overhead ratios.
Papers that use the arithmetic mean (or other averaging strategies
such as using the median) are guilty
of benchmarking crime \emph{B5: incorrect averaging across benchmark scores}.
This benchmarking crime threatens \Prop{soundness} because it results
in reporting incorrect overall overhead numbers.

\subsection{Using the wrong benchmarks}

The next group of benchmarking crimes is about using the wrong benchmarks.
It consists of three benchmarking crimes.
\emph{C1: benchmarking of simplified simulated system} refers to cases where
the benchmarks are not run on a real system but rather an emulated version,
for example through virtualization. While it is sometimes necessary to
emulate a system if it is not available otherwise, it is best avoided because
the characteristics of the emulated system are generally not identical
to those of the real system. This results
in \Prop{unsound} measurements that do not reflect the intended system.
The second is \emph{C2: inappropriate and misleading benchmarks},
which refers to the use of benchmarks that are not suitable to measure
the expected overheads. For example, it would be inappropriate
to use a workload that mostly performs user-space computations
if overhead is expected only on system calls in the kernel.
Presenting the results from inappropriate benchmarks misleads the reader
and therefore violates the property of \Prop{relevance}.
Finally, papers commit \emph{C3: same dataset for calibration and validation}
when they benchmark their system using the same data set that they used
to train it or, more generally, if there is any overlap between the training
and test sets. A typical example would be profile-guide approaches
which optimize for a specific workload and then use (parts of) that same workload
to demonstrate the performance of the technique.
The results from this approach lack \Prop{relevance}
because they mislead the reader into believing the system performs better
than it actually would in realistic scenarios.

\subsection{Improper comparison of benchmarking results}

Raw measurements like runtime or throughput numbers are rarely meaningful
in isolation. Instead, they get meaning by comparing them to a baseline
to determine how much overhead the system incurs and/or to competing systems
to determine whether the system can improve their performance.
We separated this issue into three different benchmarking crimes.
\emph{D1: no proper baseline} refers to computing overhead compared to
an unsuitable baseline. In systems defenses, the proper baseline is usually
the original system using default settings with no defenses enabled.
If the baseline is modified, for example by adding part of the requirements
for the system being evaluated (such as specific compiler flags or
virtualization), this misleads the reader by hiding some of the overhead
in the baseline and therefore violates the \Prop{relevance} requirement.
\emph{D2: only evaluate against yourself} refers to cases where papers
compare their new system to their own earlier work rather than
the state of the art. If better solutions are available,
they should be included in the comparison so as to not mislead the reader.
In this case, the comparison is not \Prop{relevant}.
Finally, \emph{D3: unfair benchmarking of competitors} refers to papers
that do compare against competitors but do so in an unfair way.
For example, they might use a configuration that is not optimal.
Again, this misleads the reader into thinking the presented system
is better than it actually is and violates \Prop{relevance}.

\subsection{Benchmarking omissions}

This group covers a number of necessary measurements for evaluations that are
not yet covered by the other benchmarking crimes.

\emph{E1: not all contributions evaluated} refers to cases where
a paper claims to achieve a certain goal, but does not empirically determine
whether this goal has been reached. It is critical that papers verify claims
for the progress of science, since incorrect claims may prevent later work that
does make the contributions from being published.
This crime violates \Prop{completeness}.

When evaluating their performance, many papers measure runtime overhead.
However, there are often other types of overhead that are also relevant
for performance. A typical example would be memory overhead.
Memory is a limited resource, so applications using lots of memory can slow down
other processes running on the same system.
Since most defenses need to use memory for bookkeeping, it is important
to measure memory consumption. A paper commits benchmarking crime
\emph{E2: only measure runtime overhead} and its evaluation is
\Prop{incomplete} whenever it does not measure
important performance characteristics.

Many systems defenses monitor behavior to determine whether it is benign or
could be malicious, which is usually impossible to do with certainty.
Unless it is obvious that the system can never get it wrong (e.g.,
security enforcement based on conservative program analysis),
the evaluation needs to quantify such failures; omission of this
assessment results in
benchmarking crime \emph{E3: false positives/negatives not tested}.
Without knowing how accurate the system is, it is impossible to tell
how valuable it would be in practice and the paper is \Prop{incomplete}.

Many systems consist of multiple components or steps that can to some extent
be used independently. For example, an instrumentation-based system might use
static analysis to eliminate irrelevant instrumentation points and improve
performance. Such optimizations are optional as they do not affect functionality
and can greatly increase complexity, so it is best to only include them if they
result in substantial performance gains. Papers that do not measure the impact
of such optional components individually commit benchmarking crime
\emph{E4: elements of solution not tested incrementally}
and its evaluation lacks \Prop{completeness}.
This is of particular importance if the optional components are a major part
of the paper's contributions. If the system is faster than the state of the art
merely due to a faster implementation rather than the newly designed
optimizations, its novelty is questionable.

\subsection{Missing information}

The final group contains benchmarking crimes where important information has
been left out of a paper.
A paper commits \emph{F1: missing platform specification} if it lacks
a description of the hardware setup used to perform the
experiments.
To be able to reproduce the results, it is always important to know
what type of CPU was used and how much memory was available.
The cache architecture may be important to understand some performance effects.
Depending on the type of system being evaluated,
other characteristics such as hard drives and networking setup may also
be essential for \Prop{reproducibility}.
The second crime in this group, \emph{F2: missing software versions},
is similar but refers to the software. It is almost always important
to specify the type and version of operating system used,
while other information such as hypervisors or compiler versions
is also commonly needed. Like the previous crime,
such omissions lead to a lack of \Prop{reproducibility}.
Next \emph{F3: subbenchmarks not listed} applies to papers that run
a benchmarking suite but do not present the results of the individual
subbenchmarks, just the overall number. This threatens \Prop{completeness} as
the results on subbenchmarks often carry important information
about the strong and the weak points of the system.
Moreover, it is important to know whether the overhead is consistent
across different applications or highly application-specific.
Finally, papers commit \emph{F4: relative numbers only} if they present
only ratios of overheads (example: system X has half the overhead of system Y)
without presenting the overhead itself (example: system X incurs 10\% overhead).
This is a bad crime as the most important result is withheld and
the reader cannot perform a sanity check of whether the results seem reasonable,
threatening the evaluation's \Prop{completeness}.
A weaker version of this practice---presenting
overheads compared to a baseline without presenting absolute runtimes
or throughput numbers---is also undesirable. The absolute numbers are valuable
for the reader to perform a sanity check (is the system configured
in a reasonable way?) and because a slow baseline often means overhead
will be less visible.
The practice of omitting absolute numbers is not harmful enough to
consider it a benchmarking crime, but we do strongly encourage authors to
include absolute numbers in addition to overheads.

%% file: tables/impact.tex
\begin{tabular}{ll|cccc|}
\multicolumn{1}{l}{} &
\multicolumn{1}{l|}{} &
\multicolumn{1}{l}{Completeness} &
\multicolumn{1}{l}{Relevancy} &
\multicolumn{1}{l}{Soundness} &
\multicolumn{1}{l|}{Reproducibility} \\
\hline
A1 & Not evaluating potential performance degradation          & $\CIRCLE$    &           &           &                 \\
A2 & Benchmark subsetting without proper justification         & $\ocircle$   & $\ocircle$&           &                 \\
A3 & Selective data set hiding deficiencies                    & $\ocircle$   &           &           &                 \\
B1 & Microbenchmarks representing overall performance          &              & $\ocircle$&           &                 \\
B2 & Throughput degraded by x\% \(\Rightarrow\) overhead is x\%&              &           & $\CIRCLE$ &                 \\
B3 & Creative overhead accounting                              &              &           & $\ocircle$&                 \\
B4 & No indication of significance of data                     & $\ocircle$   &           &           &                 \\
B5 & Incorrect averaging across benchmark scores               &              &           & $\ocircle$&                 \\
C1 & Benchmarking of simplified simulated system               &              &           & $\CIRCLE$ &                 \\
C2 & Inappropriate and misleading benchmarks                   &              & $\CIRCLE$ &           &                 \\
C3 & Same dataset for calibration and validation               &              & $\ocircle$&           &                 \\
D1 & No proper baseline                                        &              & $\CIRCLE$ &           &                 \\
D2 & Only evaluate against yourself                            &              & $\ocircle$&           &                 \\
D3 & Unfair benchmarking of competitors                        &              & $\CIRCLE$ &           &                 \\
E1 & Not all contributions evaluated                           & $\CIRCLE$    &           &           &                 \\
E2 & Only measure runtime overhead                             & $\ocircle$   &           &           &                 \\
E3 & False positives/negatives not tested                      & $\ocircle$   &           &           &                 \\
E4 & Elements of solution not tested incrementally             & $\ocircle$   &           &           &                 \\
F1 & Missing platform specification                            &              &           &           & $\ocircle$      \\
F2 & Missing software versions                                 &              &           &           & $\ocircle$      \\
F3 & Subbenchmarks not listed                                  & $\CIRCLE$    &           &           &                 \\
F4 & Relative numbers only                                     & $\ocircle$   &           &           &                 \\
\hline
\end{tabular}

%% file: sections/methodology.tex
\section{Methodology}

To determine the prevalence of the benchmarking crimes discussed
in Section~\ref{sec:crimes} and get a better idea of how papers
commit these crimes in practice, we performed a survey of 50 papers
published at top systems security venues.
Table~\ref{tab:papers} presents an overview of all the papers selected
for our analysis, sorted by year and title.

\begin{table*}[t]
\centering
\caption{Papers selected for inclusion in our analysis, sorted by year and
title\label{tab:papers}.}
\input{tables/papers.tex}
\end{table*}

Given our focus on systems security, our methodology is based on the approaches
used in prior large-scale surveys of papers in the area of
computer systems~\cite{rossow2012prudent,kurkowski2005manet,traeger2008nine,mytkowicz2009producing}.
In this section, we discuss how we performed the survey.
First we consider how to determine whether a given paper has committed
a given crime, next we discuss how we selected top venues to survey papers from,
and finally we present the sample of papers that we selected and
the rationale behind this selection.

\subsection{Classification methodology}

Based on the criteria discussed in Section~\ref{sec:crimes},
two persons independently categorized each paper for each crime as correct, flawed,
underspecified, or not applicable. In most cases, both readers came to the same 
conclusions, showing that our methodology is highly reproducible. For papers where there were some disagreements, the readers discussed
their assessments to converge on a final classification. This was the case for 8 out of 50 papers (16\%).
In only two cases did the discussion lead to the addition of a benchmarking crime that was initially missed by one of the readers.
Only one of these cases concerned a high-impact benchmarking crime.
The remaining disagreements concerned the precise extent of crimes identified by both readers, rather than crimes completely missed by one of the readers.

We use only information from the papers themselves and did not contact the
authors for explanation. Effectively, we impose on ourselves the same constraints
reviewers face when deciding whether to accept or reject a paper
in a double-blind submission system.
In cases where the papers themselves were excessively unclear about the procedures
that led to the presented results, we classified that paper/crime pair
as underspecified. This hampers reproducibility, which is a problem in itself.
We discuss this as a separate verdict in Section~\ref{sec:survey}.
Not applicable refers to cases where a paper could not possibly
have committed the crime, such as invalid comparisons in a paper
that does not perform any comparisons.

While we make an effort to anonymize our survey and prevent naming and
shaming of the papers in our sample, we do believe it is important
to allow reproduction of this study.
For this reason, we include a full overview of all evaluated papers, as well as
our detailed reasoning behind the classification of
all borderline cases (without identifying the papers themselves),
in Appendix~\ref{sec:surveyappendix}.

\subsection{Selected venues}

We focused our analysis on the traditional ``top~4'' venues in systems
security: USENIX Security, Security \& Privacy, CCS,
and NDSS. While there are many other lower-tier venues publishing
relevant systems security research, the ``top~4'' venues are
the most influential and de-facto set the standard
for benchmarking practices in the community. For our purposes,
we selected all the relevant papers from these venues in 2010
and 2015. The 2015 sample is useful to study recent trends.
The 2010 sample, in turn, allows us to examine the evolution
of benchmarking crimes over time and the impact
of Heiser's original benchmarking crimes web page~\cite{gernot}
in the systems security community five years after its publication.

\subsection{Selected papers}

From the listed conferences, we selected systems defense papers
given the increasingly strong focus on practical defense solutions in the community.
When evaluating these solutions, it is crucial to follow adequate
benchmarking practices to demonstrate that the proposed
design point in the performance-security space
actually improves the state of the art.

Among many security defense papers, it is important to clearly delimit which
papers are included and which are not to ensure reproducibility. We want to
select a group of papers for which runtime performance is of particular
importance and which are reasonably comparable among each other.
For this reason, we specifically focus on systems intended to defend software
against attacks at runtime in production settings.
For example we include sandboxing approaches,
which can be used in production to limit the damage an attacker can do,
but exclude taint tracking, which, in modern practical defenses,
is primarily used only for offline analysis.
Moreover, we only consider systems that should be expected to have
a potential runtime performance impact.
We consider approaches that modify existing software rather
than building completely new software, which allows overhead to be computed
relative to the original software baseline.

As expected, the defense
papers selected according to our criteria have a relevant presence in all the
``top~4'' venues. The steep increase of papers in 2015 (34 vs. 16 in 2010)
stands out, confirming that the number of practical defense papers and
thus the relevance of benchmarking crimes in our community is on the rise.

%% file: tables/papers.tex
\begin{tabular}{lrll}
\multicolumn{1}{l}{venue} &
\multicolumn{1}{l}{year} &
\multicolumn{1}{l}{authors} &
\multicolumn{1}{l}{title} \\
\hline
USENIX Sec & 2010 & Sehr et al.         & Adapting Software Fault Isolation to Contemporary CPU [\dots{}]                           \\
USENIX Sec & 2010 & Ter Louw et al.     & AdJail: Practical Enforcement of Confidentiality and Integrity [\dots{}]                  \\
CCS        & 2010 & Lu et al.           & BLADE: An Attack-Agnostic Approach for Preventing [\dots{}]                               \\
USENIX Sec & 2010 & Watson et al.       & Capsicum: Practical Capabilities for UNIX                                                 \\
USENIX Sec & 2010 & Akritidis           & Cling: A Memory Allocator to Mitigate Dangling Pointers                                   \\
S\&P       & 2010 & Meyerovich et al.   & ConScript: Specifying and Enforcing Fine-Grained Security [\dots{}]                       \\
CCS        & 2010 & Novark et al.       & DieHarder: Securing the Heap                                                              \\
S\&P       & 2010 & Wang                & HyperSafe: A Lightweight Approach to Provide Lifetime [\dots{}]                           \\
CCS        & 2010 & Azab et al.         & HyperSentry: Enabling Stealthy In-context Measurement of [\dots{}]                        \\
NDSS       & 2010 & Seo et al.          & InvisiType: Object-Oriented Security Policies                                             \\
USENIX Sec & 2010 & Kim et al.          & Making Linux Protection Mechanisms Egalitarian with UserFS                                \\
S\&P       & 2010 & Devriese et al.     & Non-Interference Through Secure Multi-Execution                                           \\
CCS        & 2010 & Askarov et al.      & Predictive Black-box Mitigation of Timing Channels                                        \\
NDSS       & 2010 & Barth et al.        & Protecting Browsers from Extension Vulnerabilities                                        \\
CCS        & 2010 & Cappos et al.       & Retaining Sandbox Containment Despite Bugs in Privileged [\dots{}]                        \\
USENIX Sec & 2010 & Djeric et al.       & Securing Script-Based Extensibility in Web Browsers                                       \\
CCS        & 2015 & Lu et al.           & ASLR-Guard: Stopping Address Space Leakage for Code Reuse [\dots{}]                       \\
USENIX Sec & 2015 & Backes et al.       & Boxify: Full-fledged App Sandboxing for Stock Android                                     \\
CCS        & 2015 & Mashtizadeh et al.  & CCFI: Cryptographically Enforced Control Flow Integrity                                   \\
USENIX Sec & 2015 & Araujo et al.       & Compiler-instrumented, Dynamic Secret-Redaction of Legacy [\dots{}]                       \\
NDSS       & 2015 & Song et al.         & Exploiting and Protecting Dynamic Code Generation                                         \\
CCS        & 2015 & Muthukumaran et al. & FlowWatcher: Defending against Data Disclosure [\dots{}]                                  \\
NDSS       & 2015 & Younan              & FreeSentry: protecting against use-after-free [\dots{}]                                   \\
CCS        & 2015 & Tang et al.         & Heisenbyte: Thwarting Memory Disclosure Attacks using [\dots{}]                           \\
S\&P       & 2015 & Wagner et al.       & High System-Code Security with Low Overhead                                               \\
NDSS       & 2015 & Davi et al.         & Isomeron: Code Randomization Resilient to (Just-In-Time) [\dots{}]                        \\
CCS        & 2015 & Chudnov et al.      & Inlined Information Flow Monitoring for JavaScript                                        \\
CCS        & 2015 & Crane et al.        & It's a TRaP: Table Randomization and Protection against [\dots{}]                         \\
S\&P       & 2015 & Zhang et al.        & Leave Me Alone: App-level Protection Against Runtime [\dots{}]                            \\
USENIX Sec & 2015 & Feng et al.         & LinkDroid: Reducing Unregulated Aggregation of App Usage [\dots{}]                        \\
NDSS       & 2015 & Mohan er al.        & Opaque Control-Flow Integrity                                                             \\
CCS        & 2015 & Niu et al.          & Per-Input Control-Flow Integrity                                                          \\
CCS        & 2015 & Van der Veen et al. & Practical Context-Sensitive CFI                                                           \\
NDSS       & 2015 & Lee et al.          & Preventing Use-after-free with Dangling Pointers Nullification                            \\
S\&P       & 2015 & Guan et al.         & Protecting Private Keys against Memory Disclosure Attacks [\dots{}]                       \\
USENIX Sec & 2015 & Rane et al.         & Raccoon: Closing Digital Side-Channels through Obfuscated [\dots{}]                       \\
S\&P       & 2015 & Stephen et al.      & Readactor: Practical Code Randomization Resilient to Memory [\dots{}]                     \\
NDSS       & 2015 & Jang et al.         & SeCReT: Secure Channel between Rich Execution Environment [\dots{}]                       \\
NDSS       & 2015 & Chen et al.         & StackArmor: Comprehensive Protection From Stack-based [\dots{}]                           \\
CCS        & 2015 & Soni et al.         & The SICILIAN Defense: Signature-based Whitelisting of Web [\dots{}]                       \\
NDSS       & 2015 & Crane et al.        & Thwarting Cache Side-Channel Attacks Through Dynamic [\dots{}]                            \\
CCS        & 2015 & Liu et al.          & Thwarting Memory Disclosure with Efficient [\dots{}]                                      \\
CCS        & 2015 & Bigelow et al.      & Timely Rerandomization for Mitigating Memory Disclosures                                  \\
USENIX Sec & 2015 & Lee et al.          & Type Casting Verification: Stopping an Emerging Attack Vector                             \\
CCS        & 2015 & Xu et al.           & UCognito: Private Browsing without Tears                                                  \\
S\&P       & 2015 & Schuster et al.     & VC3: Trustworthy Data Analytics in the Cloud using SGX                                    \\
NDSS       & 2015 & Prakash et al.      & vfGuard: Strict Protection for Virtual Function Calls in COTS [\dots{}]                   \\
NDSS       & 2015 & Zhang et al.        & VTint: Protecting Virtual Function Tables' Integrity                                      \\
NDSS       & 2015 & Demetriou et al.    & What's in Your Dongle and Bank Account? Mandatory and [\dots{}]                           \\
USENIX Sec & 2015 & Weissbacher et al.  & ZigZag: Automatically Hardening Web Applications Against [\dots{}]                        \\
\hline
\end{tabular}

%% file: sections/survey.tex
\section{Survey results}
\label{sec:survey}

For each selected paper listed in Table~\ref{tab:papers} and
each benchmarking crime described in Section~\ref{sec:crimes},
we have determined whether the paper commits that particular benchmarking crime.
Table~\ref{tab:crimes} provides the number of papers committing each crime
split by year of publication. In this table, we consider only whether
the paper commits the crime at least once (i.e., papers that commit the same
benchmarking crime multiple times are counted once).
In some cases, we were unable to determine whether the methodology
in the paper is sound because important elements of the experiments or
their analysis were not specified with a sufficient level of detail.
We have classified these paper/crime pairs as underspecified.
It is important to note that underspecification is problematic
even if the underlying methodology is sound as it hampers reproducibility
and makes it harder for later competitors to perform a fair comparison
with prior work.

\begin{table*}[t]
\centering
\caption{Benchmarking crimes survey overview\label{tab:crimes}.}
\resizebox{6.5in}{!}{
\input{tables/crimes.tex}
}
\end{table*}

Our results show that benchmarking crimes are a major problem in both years
we investigate. Over all pairs of a paper and an applicable crime,
the crime either applies or the paper is underspecified with regard
to the crime in 77 out of the 255 cases (30\%) for 2010 and
in 179 out of the 596 cases (30\%) for 2015.
However, not all crimes are equally common.
The lack of indication of significance of data and
benchmark subsetting without proper justification are
by far the most widespread, respectively affecting 80\% and 69\%
of the applicable papers we surveyed. None of the other crimes
affect a majority of the papers, but four additional ones affect 40\%
or more of the papers to which they apply.
This shows that several types of benchmarking crimes are widespread
even in peer-reviewed papers at top venues.

There is no clear difference visible between the more recent and
the older papers, confirming that improper benchmarking is
a longstanding problem and that the original web page
on benchmarking crimes published in 2010~\cite{gernot}
did not have a sufficient impact in the systems security community.
The fraction
of paper/crime pairs that applies or is underspecified is almost identical
between the years (30\% in 2010, 30\% in 2015). 

For most individual benchmarking crimes we cannot apply the $\chi^2$-test
directly because the expected values in some cells
are below 5~\cite{yates1934contingency}.
This is mostly due to the fact that there were relatively
few suitable papers published in 2010.
In the cases where the $\chi^2$-test does (almost) apply,
the differences between the years are always insignificant.
This is the case for benchmarking crimes A1, B2, and E2
(see Table~\ref{tab:crimes} for the numbering).
In the other cases, we apply Yates' correction
for continuity~\cite{yates1934contingency} and find significant differences
only for benchmarking crime E1 ($p=0.001$).
The number of papers in which
not all contributions are evaluated (crime E1) has gone down significantly
over our period of five years, which suggests that either authors or
reviewers have been more careful to require a complete evaluation.
Overall, however, our conclusion must be that differences over time
are minor and, in almost all cases, statistically insignificant for our sample.

Based on our findings in the survey, we classified some benchmarking crimes
as \emph{high-impact} to indicate that they are almost always a major threat
to the usefulness of the evaluation and, with it, the scientific value
of the paper. Table~\ref{tab:impact} presents our classification.
We discuss the concrete impact for each individual crime
in Section~\ref{sec:impact}.
A typical example of a high-impact crime is not evaluating
all contributions, as unverified claims cannot be considered true contributions.
A typical example of a crime that is not high-impact is using the arithmetic
mean to average overhead numbers; while the impact is severe in specific cases,
there are also papers where the difference is small and therefore does not
undermine the value of the paper. While we recognize any such classification
is necessarily subjective, we did make an effort to reflect our observations
from the survey. We do believe that any high-impact crime we listed should be
a reason for reviewers to demand the paper to be revised,
while for the other crimes this depends on the context.
Overall, high-impact crimes are somewhat less common than other crimes.
In our sample we found 86 high-impact crimes out of 346 applicable
crime/paper pairs (25\%) and 167 other crimes out of 505 applicable pairs
(33\%). A $\chi^2$ test shows this difference to be significant with $p<0.0005$.

\begin{figure}[!t]
\centering
\includegraphics[width=3in,trim={3.0cm 3.0cm 2.6cm 2.5cm}]{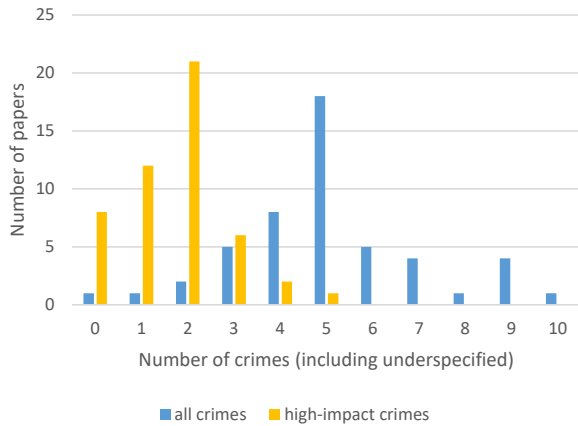}
\caption{Histogram of number of crimes per
  paper\label{fig:crimes-per-paper}}
\end{figure}

Figure~\ref{fig:crimes-per-paper} shows a histogram of the number
of benchmarking crimes (including underspecification) per paper.
It is notable that from our sample of 50 papers,
we found only a single paper without any benchmarking crimes.
Crimes are fairly evenly spread between papers,
with many papers being very close to the average number of
benchmarking crimes per paper (5.0 for all crimes, 1.7 for high-impact crimes).
As such, the results would seem
to suggest that the problem of benchmarking crimes is not an issue of
a few authors and reviewers being particularly careless (or malicious),
but rather a community-wide lack of awareness of or attention to
these problems. This is further corroborated by the fact
that many prevalent benchmarking crimes require very little
effort to fix, as detailed later.

For completeness and to improve transparency, we have included a detailed
discussion and justification of the way we classified the papers
in Appendix~\ref{sec:surveyappendix}.

%% file: tables/crimes.tex
\begin{tabular}{ll|rrlrl|rrlrl|}
\multicolumn{2}{l|}{} &
\multicolumn{5}{l|}{2010} &
\multicolumn{5}{l|}{2015} \\
\multicolumn{2}{l|}{} &
\multicolumn{1}{l}{appl.} &
\multicolumn{2}{l}{flawed} &
\multicolumn{2}{l|}{undersp.} &
\multicolumn{1}{l}{appl.} &
\multicolumn{2}{l}{flawed} &
\multicolumn{2}{l|}{undersp.} \\
\hline
A1 & Not evaluating potential perf. degradation                &         16 &      8 & (50\%) &              0 & (0\%)  &         34 &      8 & (24\%) &              1 & (3\%)  \\
A2 & Benchmark subsetting w/o proper justification             &          9 &      4 & (44\%) &              0 & (0\%)  &         33 &     24 & (73\%) &              1 & (3\%)  \\
A3 & Selective data set hiding deficiencies                    &         16 &      1 & (6\%)  &              1 & (6\%)  &         32 &      6 & (19\%) &              0 & (0\%)  \\
B1 & Microbenchmarks representing overall perf.                &         14 &      5 & (36\%) &              0 & (0\%)  &         10 &      1 & (10\%) &              0 & (0\%)  \\
B2 & Throughput degr. by x\% \(\Rightarrow\) overhead is x\%   &         13 &      6 & (46\%) &              2 & (15\%) &         30 &     10 & (33\%) &              0 & (0\%)  \\
B3 & Creative overhead accounting                              &         16 &      1 & (6\%)  &              1 & (6\%)  &         34 &      8 & (24\%) &              1 & (3\%)  \\
B4 & No indication of significance of data                     &         16 &     13 & (81\%) &              0 & (0\%)  &         34 &     25 & (74\%) &              2 & (6\%)  \\
B5 & Incorrect averaging across benchmark scores               &          5 &      0 & (0\%)  &              2 & (40\%) &         24 &     12 & (50\%) &              0 & (0\%)  \\
C1 & Benchmarking of simplified simulated system               &         16 &      2 & (13\%) &              0 & (0\%)  &         34 &      3 & (9\%)  &              0 & (0\%)  \\
C2 & Inappropriate and misleading benchmarks                   &         16 &      0 & (0\%)  &              1 & (6\%)  &         34 &      8 & (24\%) &              1 & (3\%)  \\
C3 & Same dataset for calibration and validation               &          0 &      0 &        &              0 &        &          5 &      1 & (20\%) &              3 & (60\%) \\
D1 & No proper baseline                                        &         16 &      3 & (19\%) &              0 & (0\%)  &         34 &      9 & (26\%) &              5 & (15\%) \\
D2 & Only evaluate against yourself                            &          2 &      0 & (0\%)  &              0 & (0\%)  &         13 &      2 & (15\%) &              0 & (0\%)  \\
D3 & Unfair benchmarking of competitors                        &          2 &      0 & (0\%)  &              1 & (50\%) &         13 &      4 & (31\%) &              1 & (8\%)  \\
E1 & Not all contributions evaluated                           &         16 &      6 & (38\%) &              0 & (0\%)  &         34 &      0 & (0\%)  &              0 & (0\%)  \\
E2 & Only measure runtime overhead                             &         16 &      6 & (38\%) &              0 & (0\%)  &         34 &     17 & (50\%) &              0 & (0\%)  \\
E3 & False positives/negatives not tested                      &          5 &      3 & (60\%) &              0 & (0\%)  &         14 &      3 & (21\%) &              0 & (0\%)  \\
E4 & Elements of solution not tested incrementally             &          5 &      0 & (0\%)  &              0 & (0\%)  &         20 &      4 & (20\%) &              0 & (0\%)  \\
F1 & Missing platform specification                            &         16 &      4 & (25\%) &              0 & (0\%)  &         34 &      7 & (21\%) &              0 & (0\%)  \\
F2 & Missing software versions                                 &         16 &      5 & (31\%) &              0 & (0\%)  &         34 &      7 & (21\%) &              0 & (0\%)  \\
F3 & Subbenchmarks not listed                                  &          8 &      2 & (25\%) &              0 & (0\%)  &         30 &      5 & (17\%) &              0 & (0\%)  \\
F4 & Relative numbers only                                     &         16 &      0 & (0\%)  &              0 & (0\%)  &         32 &      0 & (0\%)  &              0 & (0\%)  \\
\hline
   & Total                                                     &        255 &     69 & (27\%) &              8 & (3\%)  &        596 &    162 & (27\%) &             15 & (3\%)  \\
\hline
\end{tabular}

%% file: sections/impact.tex
\section{Impact}
\label{sec:impact}

In this section, we consider the impact of the various benchmarking crimes
based on our findings from the survey we conducted.

\subsection{Selective benchmarking}

\paragraph{A1 - Not evaluating potential performance degradation}
We found two major groups of papers that commit this crime:
those where overhead figures are missing entirely and
those that do not reflect all potential slowdown.
In both cases, this crime makes it difficult (if not impossible)
to assess the practicality of the presented solution and improvements
over the state of the art. Moreover, papers that present inappropriate
performance measurements may even hamper scientific progress because they
prevent competing systems that perform poorly on these inappropriate measures or
not as efficiently on appropriate measures from being published. Even worse,
they may encourage more benchmarking crimes in future systems,
as authors struggle to beat overly optimistic performance figures.
As such, we consider this crime high-impact.

\paragraph{A2 - Benchmark subsetting without proper justification}
We found that many papers that use standardized benchmarking suites
leave out some subbenchmarks. Based on the particular benchmarks that are
often left out, it is very likely that this will result in an underestimate
of performance in practice (see Appendix~\ref{sec:surveyappendix-selective}
for details). We conclude that leaving out subbenchmarks can have a major impact
on the soundness of measurements as well as the comparability
between competing systems and therefore requires a proper and
explicit justification.
Moreover, if different papers use subsets, the overall slowdown is
no longer suitable for comparing performance.
Fortunately, many of these problems can be solved simply by explicitly
acknowledging that a paper uses a subset of the available subbenchmarks
and detailing the reasoning behind this choice.
Despite the possibly large impact we do not consider this crime
to be high-impact as there are also cases where the particular subbenchmarks
left out do not seem to introduce a bias.

\paragraph{A3 - Selective data set hiding deficiencies}
We found two types of occurrences of this benchmarking crime,
with different impacts.
Papers with important missing variables make it hard to estimate
how the solution would behave in practical situations and
may hide limitations of the solution's performance.
Papers which use variables with a restricted range might result
in incorrect extrapolation and again hide limitations.

\subsection{Improper handling of benchmark results}

\paragraph{B1 - Microbenchmarks representing overall performance}

This crime came in two flavors in our survey:
papers which leave out macrobenchmarks altogether and
one paper that includes both but bases performance claims on microbenchmarks.
In both cases this is inappropriate as microbenchmarks
are a poor indicator for real-world performance, resulting in misleading claims.
In the former case it is impossible to determine how strong this impact is,
but in the latter case the paper suggested a runtime performance
that is not realistic in practice.

\paragraph{B2 - Throughput degraded by x\% \(\Rightarrow\) overhead is x\%}

Based on our survey, we believe that all instances of this benchmarking crime
are likely to result in an underestimate of performance overhead,
although without the necessary data it is impossible to determine by how much.
Because this crime is likely to affect the soundness of performance
measurements in all cases, we consider it to be a high-impact
benchmarking crime.

\paragraph{B3 - Creative overhead accounting}

It is hard to make a general statement about the impact of creative overhead
accounting as this benchmarking crime is committed in many different ways.
In some papers this leads to unsound results, some of which systematic
underestimations of overhead, while in other cases the conclusions
are misleading. See Appendix~\ref{sec:surveyappendix-selective}
for details about the specific issues we found.

\paragraph{B4 - No indication of significance of data}

Some indication of variation is important because it is an indication
of how reliable the numbers are and whether, given the measurement inaccuracy,
the measured differences are actually meaningful.
However, we expect the overall impact of this crime to be relatively mild
for papers where researchers set up their experiments correctly.

\paragraph{B5 - Incorrect averaging across benchmark scores}

To determine the impact of incorrect averaging, we computed the geometric mean
based on tables or graphs presenting the subbenchmark results
for papers that should have used it.
Because there is some inaccuracy in deriving numbers from the graphs,
we compared the geometric mean with the arithmetic mean derived from the same
numbers rather than the arithmetic mean presented in the paper.
We were able to do this for eight papers.
For four out of the eight papers, the difference between the means is less than
1\% and as such the impact of using the incorrect mean is negligible.
For the other four papers, the arithmetic mean is higher than
the geometric mean, so they overestimate overall overhead.
In the worst case we found, the arithmetic mean is more than twice
the geometric mean, while the remainder overestimates overhead by 2\% to 16\%.
The relative difference between the means is largest in cases
where the overhead is large.

\subsection{Using the wrong benchmarks}

\paragraph{C1 - Benchmarking of simplified simulated system}

For all papers that committed this benchmarking crime,
benchmarking a simplified system threatens the accuracy
of the reported numbers and makes it harder to compare against competing
systems that were evaluated under more realistic conditions.
Given that this issue always yields potentially unsound results,
we classified it as high-impact.

\paragraph{C2 - Inappropriate and misleading benchmarks}

In all cases we found, the use of inappropriate and misleading benchmarks
is likely to have a major impact on the validity of the results.
This crime always results in either an underestimate of overhead or
an overestimate of effectiveness in the papers in our survey.
For this reason, we consider this a high-impact crime.

\paragraph{C3: same dataset for calibration and validation}

While we believe this is a very serious crime that can have a major impact,
we have found too few papers that it applies to in our sample to meaningfully
judge its impact in practice. However, we believe that as profile-guiding
and machine learning become more popular, this may become a major issue
if authors and reviewers do not pay sufficient attention to it.

\subsection{Improper comparison of benchmarking results}

\paragraph{D1 - No proper baseline}

With regard to the impact, we can distinguish two different cases
for this crime: papers that have an incorrect baseline and papers
that do not present one at all.
In our sample, the former group is always likely to either
underestimate overhead or overestimate effectiveness.
This threatens both the soundness and comparability of the results.
Absolute performance numbers with no baseline to compare against
cannot be compared between systems and therefore provide little meaningful
information. Since we found that the lack of a proper baseline was a serious
problem in all cases, we consider this crime high-impact.

\paragraph{D2 - Only evaluate against yourself}

The impact for this crime in practice is hard to assess because it would
require gathering the state of the art at the time the paper was submitted
for publication and ensuring their performance numbers are actually comparable.
This process would be highly error-prone except when done
by an expert on the type of system the paper is about.

\paragraph{D3 - Unfair benchmarking of competitors}

In all cases of this benchmarking crime that we found, the reader is misled
into believing the presented system performs better compared
to the state of the art than it actually does.
As such, we consider this crime to be high-impact.

\subsection{Benchmarking omissions}

\paragraph{E1 - Not all contributions evaluated}

The impact of not evaluating claimed contributions is that
the design may not actually work as advertised and future
solutions that do achieve such goals may have a much harder time
getting published, holding back research progress.
Given that this risk is present in all cases we found,
this crime is labeled as high-impact.

\paragraph{E2 - Only measure runtime overhead}

Papers which do not measure important sources of overhead other than runtime
are incomplete. However, the impact of this incompleteness differs
from case to case. If, for example, memory overhead can theoretically
be assumed to be minor and similar to prior work, the impact is limited.
If, on the other hand, there is reason to believe the paper incurs significant
memory overhead yet does not measure it, this could be a problem for later
papers that improve on this overhead.

\paragraph{E3 - False positives/negatives not tested}

The lack of testing for false positives or negatives is potentially
a major issue because if the number of these is substantial it could greatly affect
the practicality or effectiveness of the approach. Without this information,
it may be impossible for a reader to properly assess how valuable
the contributions of the paper are. That said, in practice the impact depends
on the type of system presented. In some cases false positives may crash
the system while in others they merely result in performance degradation.

\paragraph{E4 - Elements of solution not tested incrementally}

If elements of the presented system are not tested incrementally,
it is unclear whether all parts of the approach are
indeed necessary to implement a system that is as effective and efficient
and therefore it is also unclear whether all the components are actually
contributions.

\subsection{Missing information}

\paragraph{F1 - Missing platform specification}

In all cases, this benchmarking crime makes reproducing the exact results
based on the contents of the paper impossible
and it may make the results less comparable.
However, it does not affect the validity of the results.

\paragraph{F2 - Missing software versions}

This benchmarking crime hampers reproducibility for all papers that commit it,
as the software about which information is missing should be expected
to have an impact on performance.

\paragraph{F3 - Subbenchmarks not listed}

The impact of this benchmarking crime is somewhat hard to estimate.
Although the lack of important information always affects completeness
of the paper, it may even result in measurements that are unsound
and misleading. This is the case, for example, if the omission obscures the fact
that the results are greatly affected by outliers or that only a subset of
the benchmarking suite is run. The latter also makes the results incomparable.
While it is impossible to tell whether this is the case due to the missing
information, our results for crime A2 suggest the practice of unjustified
subsetting is widespread. Because of the wide range of possible consequences
of this crime it seems likely there is some relevant impact for almost
every paper that commits this crime and, as such, we consider it high-impact.

\paragraph{F4 - Relative overheads only}

We have not found this crime in its worst form, so we cannot determine
the practical impact. As for leaving out an absolute baseline,
we have found one case of D1 (no proper baseline) where the presented
absolute baseline was clearly inconsistent with the reference baseline
for the benchmark. This means the measurement was performed incorrectly,
something that would not have been clear without the absolute baseline.
As such we believe that the mild version of this crime does impact some cases.

%% file: sections/recommendations.tex
\section{Recommendations}

While our analysis shows that benchmarking crimes are very common and
potentially have a major impact on the quality of published research
in systems security, it also reveals that the quality of published
research could be greatly improved with little effort
by paying extra attention to the most important crimes.

The primary focus should be on preventing common high-impact benchmarking crimes.
The most common high-impact benchmarking crimes are
A1 (not evaluating potential performance degradation),
B2 (throughput degraded by x\% \(\Rightarrow\) overhead is x\%), and
D1 (no proper baseline). We believe authors should consider these crimes
early on in the research process to ensure they set up the right
benchmarks.

To address A1, authors should consider which performance dimensions the solution
could possibly affect (for example, CPU, concurrency, memory, IO, system calls, \dots)
and include at least one appropriate benchmark for each dimension.
Authors can address B2 by ensuring the system is always fully loaded
while benchmarking. Typically, this is simply a matter of setting up
a sufficient number of concurrent operations on workloads that would otherwise
be bound by IO latencies. If this is not feasible, an alternative is to present
the CPU load on both the baseline and the experimental setup in the paper.
Benchmarking crime D1 can be addressed by considering the way the system
protected by the provided solution would be used in a setting
where the presented solution is not available.
Often, this means avoiding any non-default compiler flags or emulation
techniques that would slow down the baseline.
Moreover, authors should always specify what the baseline is.

A number of common benchmarking crimes is not necessarily high-impact,
but very easy to address and we believe every author should go through
the list to avoid them.
In particular, crimes B4 (no indication of significance of data),
B5 (incorrect averaging across benchmark scores),
F2 (missing software versions), and
F1 (missing platform specification) can be addressed by simply adding
readily available data to the paper.
Yet, each of these crimes is committed by more than 10 papers in the sample.
Although F4 (relative overheads only) is not committed by the papers
in our survey in the worst form, many papers can still be improved by adding
an absolute baseline.
Addressing each of these issues should take almost no time (and space),
yet it would greatly improve many of the papers in our survey.

One more benchmarking crime is neither high-impact nor trivial to address,
but it is so common that we feel it deserves more attention
since it does have a major overall impact on the quality of research
in our field. A2 (benchmark subsetting without proper justification)
does not always have a large impact, but it may result
in overly optimistic (or completely incorrect) overall overheads.
Authors should run all subbenchmarks that can reasonably be run
and be explicit about reasons for omitting the others.
Moreover, they should not present the overall result
as if it is a complete result that can be compared
with other papers using the same benchmarks.

While we hope authors avoid all the benchmarking crimes discussed
in this paper, we believe that following the recommendations here would
at least be a first step to greatly improve the research quality
in systems security with relatively little effort.
Had all the authors followed these simple rules,
it would almost triple the number of papers without any high-impact crimes
committed or underspecified (from 8 to 22 papers),
greatly increase the number of papers that commit no crimes at all
(from 1 to 9 papers), and reduce the average number of crimes per paper
by almost two thirds (4.6 to 1.7 for all crimes,
1.5 to 0.6 for high-profile crimes).

%% file: sections/limitations.tex
\section{Limitations}

Although we have performed this survey as carefully as possible,
there are a number of limitations on its applicability that are hard to avoid.

First, we do not claim that either our list of benchmarking crimes or our
dimensions of evaluation quality are complete. Similarly, we do not
seek comparison with other systems fields, as the distribution
of crimes is inherently field-specific.
There are many more benchmarking crimes possible in the broader computer systems
field. The ones we examined are merely some of the most important
issues that stand out for being common problems in systems security papers,
especially defenses.

Second, in some cases, whether a particular benchmarking crime has been
committed or even whether a benchmarking crime applies to a paper is subjective.
Other people could reach somewhat different conclusions,
although we did make an effort to be lenient in borderline cases
so as to be conservative in our analysis. We also discussed borderline cases
among ourselves and always consulted an independent reader as necessary.
Whenever possible, we explicitly discuss these cases
in Section~\ref{sec:survey}. We also cannot rule out that,
despite the care we put into our analysis, there can be mistakes or oversights.
Hopefully, this only concerns a small fraction of the paper/crime pairs.

A third limitation is the fact that we cannot be transparent about which papers
commit which crimes. While this would be better for reproducibility and
allowing others to verify our work, we believe that naming and shaming
would be counterproductive as in our opinion the problem is not with individuals
but rather the community as a whole.
Moreover, we believe it would not be appropriate to create what amounts
to a ranking of individuals or institutions given that not all crimes are
equally severe and the lack of the specific crimes we consider does not imply
that there are no other flaws in the paper.
In avoiding this, we follow common practice in papers that perform
similar surveys~\cite{rossow2012prudent,kurkowski2005manet,traeger2008nine,mytkowicz2009producing}.
To compensate, we added a detailed discussion
in Appendix~\ref{sec:surveyappendix} that should allow others to perform
the survey themselves according to the same criteria.

Fourth, popular research topics have changed over time,
which makes a direct comparison between percentages in 2010 and 2015 hard.
Different types of papers are subject to different types of benchmarking crimes.
All we can and did do is show that benchmarking crimes were a problem
at both points in time.

Fifth, for studies such as the one presented in this paper a larger sample size
is always desirable. Since we surveyed all eligible papers in all tier-1
security venues for 2010 and 2015, the most logical way to increase the sample
size would be to consider more years. However, given that we found minimal
differences between the two years currently surveyed, we believe that a larger
sample over recent years would not yield significantly different results.

Finally, published papers are not necessarily a representative sample of all
papers, especially at the top conferences.
One would hope the review process weeds out the papers
which commit the worst benchmarking crimes, but one cannot rule out
that benchmarking crimes make acceptance more likely if they are
not too obvious and appear to improve the presented results.
Either possibility creates a bias when applying our survey results
to papers submitted for review.

%% file: sections/relatedwork.tex
\section{Related Work}

\paragraph{Benchmarking in systems security}
While there have been several surveys to determine whether computer science
papers perform measurements in appropriate ways~\cite{aviv2011challenges,kurkowski2005manet,kuz2011multicore,mogul1999brittle,traeger2008nine,skadron2003challenges,mytkowicz2009producing,rossow2012prudent},
to the best of our knowledge none of them is specific to benchmarking
in systems security. 
The most closely related work is Heiser's original web page
about benchmarking crimes~\cite{gernot}, which serves as an inspiration for this
paper and forms the basis for our benchmarking requirements.
Compared to Heiser's web page, we propose an extended classification and
present a systematic analysis to show that benchmarking crimes are indeed an
increasingly relevant problem in peer-reviewed defense papers accepted at top
systems security venues. We also formulate concrete recommendations
for authors in the security community.

\paragraph{Surveys considering evaluation quality}
We will now consider a number of papers that have performed surveys
to determine how well papers in various fields evaluate their work.
Kuz et al.~\cite{kuz2011multicore} survey the use of benchmarking
for multi-core systems to propose a better approach,
but only include six papers in their survey.
Skadron et al.~\cite{skadron2003challenges} survey a number of papers
in computer architecture to determine their topics and performance evaluation
techniques adopted. They provide an overview and discussion of
the various techniques, of which benchmarking is done,
but do not go in depth about incorrect benchmarking practices.
Kurkowski et al.~\cite{kurkowski2005manet} survey papers
using simulation techniques for mobile ad-hoc networks (MANET)
and identify common pitfalls.
Krishnamurty and Willinger~\cite{krishnamurthy2008our} discuss a list of
common pitfalls in networking measurements using illustrative examples
of flaws, but do not perform a systematic survey.
Mogul~\cite{mogul1999brittle} surveys papers to determine what types
of benchmarks are commonly used in operating systems papers.
However, it considers only whether those benchmarks themselves are realistic,
not whether they are used appropriately.
Traeger and Zadok~\cite{traeger2008nine} survey the use of benchmarks
in file systems and storage research. However, they limit themselves
to setting up the benchmarks and do not consider
whether the results are handled appropriately.
Mytkowicz~\cite{mytkowicz2009producing} presents a survey to determine whether
measurement error is considered correctly in computer systems experiments
and provides suggestions on how to improve this.
Aviv and Haeberlen~\cite{aviv2011challenges} survey botnet research,
but focus on correctness evaluations rather than performance.
Collberg et al.~\cite{collberg2016repeatability} survey a number of
computer systems papers to examine their repeatability, but focus
on being able to locate, build, and run the systems prototypes.
No attempt is made to reproduce the experimental results detailed
in the paper or generally assess the quality of their benchmarking results.
Rossow et al.~\cite{rossow2012prudent} study the methodological rigor and
prudence in papers using malware execution.
While their approach to identifying flaws and surveying is similar
to ours, the pitfalls they identify are quite different
because they focus on malware analysis rather than on performance.
While these papers have used a survey of published papers in ways similar
to ours, none of them are in the field of systems security and
none considers all the benchmarking flaws we do.

\paragraph{Benchmarking advice}
Some other papers also provide benchmarking advice but do so
without a systematic survey, instead using examples, and their own tests
to verify the identified pitfalls result in questionable results.
Schwarzkopf et al.~\cite{schwarzkopf2012seven} identify benchmarking problems
in cloud research this way and Seltzer et al.~\cite{seltzer1999case}
discuss problems with using standardized benchmarks in file systems research.
While these studies demonstrate important benchmarking problems,
the lack of a survey means they cannot determine the impact these potential
problems have on the research literature in practice.

%% file: sections/conclusion.tex
\section{Conclusion}

As a security community, we struggle to preserve the integrity of everyday
systems from increasingly dangerous security threats.
Regrettably, much less attention has been devoted to preserve the integrity of
systems security research results themselves from accidental ``threats''.
Benchmarking crimes, in particular, have been largely neglected in systems
security research, as its core focus has been traditionally on security
rather than performance. As the focus of the community is increasingly shifting
to devising practical, low-overhead defenses, however,
benchmarking crimes have grown extremely relevant and are now the elephant in
the room.

In this paper, we assessed the magnitude of the problem
by surveying 50 defense papers in top systems security venues. Our results
show that benchmarking crimes are widespread and, while their prevalence
has not changed over time, their impact is increasingly
worrisome. Faults in benchmarking can hamper comparability
and reproducibility at best, or ``poison" an entire research
area in the worst case. Encouragingly, we found that many common
benchmarking crimes can be easily prevented and we formulated
concrete recommendations for authors. We hope our research will
raise awareness of this threat and encourage adequate
benchmarking practices to improve the quality of the
scientific process in our community.

%% file: sections/surveyappendix.tex
\section{Appendix: Survey details}
\label{sec:surveyappendix}

In this appendix we discuss the conclusions from our survey
for the individual benchmarking crimes introduced in Section~\ref{sec:crimes}.
In each subsection, we elaborate on one group of benchmarking crimes.
Where appropriate we use examples from the papers we surveyed,
but to keep the discussion anonymous with regard to the papers in our sample,
we either abstract away or change some of the details.
We also consider what impact the benchmarking crimes we found are likely
to have on the results.

\subsection{Selective benchmarking}
\label{sec:surveyappendix-selective}

Benchmarking crimes related to selective benchmarking are very common.
40 out of the 50 papers in our sample (80\%) commit at least one of the three
crimes in this group and one additional paper does not provide enough
information to determine whether this element is performed correctly.
This is largely due to the most common benchmarking crime in this group,
selecting a subset of a benchmarking suite without proper justification (A2).

\subsubsection*{A1 - Not evaluating potential performance degradation}

Not evaluating potential performance degradation is a relatively common
benchmarking crime, affecting 16 out of the 50 papers (32\%) it applies to,
and being underspecified in one more case.
There are two main manifestations of this crime.
The most obvious case are those papers which provide no meaningful measurement
of runtime performance for some or all of the systems presented.
We found this to be the case for seven papers in our sample.
A more subtle case are those papers that do present runtime performance numbers,
but where the benchmarks used to measure those numbers are inappropriate
for the presented system, not reflecting an important element of
its potential performance impact. This occurs for eight papers in our sample.
Examples include not using a memory-intensive benchmark for systems likely
to affect memory accesses, using a single-threaded workload for systems that
benefit from additional cores,
using benchmarks that do not stress instrumented calls,
or omitting start-up/warm-up periods that might be affected by the system.
While these papers do present runtime performance numbers,
they are not meaningful for comparisons to similar systems.

\subsubsection*{A2 - Benchmark subsetting without proper justification}

This is
the most common benchmarking crime in this group, affecting 28 out of the 42 papers
(67\%) it applies to and is underspecified in one more case.
The most common benchmarking setup in our sample is the use of
the SPEC CPU~\cite{henning2006spec} benchmarks,
which is the case for 18 out of 50 papers (36\%).
These CPU-intensive benchmarks are appropriate to test single-threaded
performance of systems that insert instrumentation which requires
the CPU and the memory to do more work to run the program.

However, many papers using SPEC CPU benchmarks only run a subset
of the benchmarks. The papers from our sample show that overhead often
differs greatly between the programs that make up the SPEC CPU
benchmarking suites, with the percentage overhead often showing
at least an order of magnitude difference between the best and the worst case.
In particular, perlbench, xalancbmk, and povray often stand out
for large overhead numbers. If any of these benchmarks is omitted,
it can have a large impact on the overall overhead computed for SPEC.
However, there is a substantial difference between the different systems
in which benchmarks stress them most, so even if other benchmarks are left
out, there can be a large and unpredictable impact on the overall result.

We find that leaving out SPEC subbenchmarks for legitimate reasons is common
and we have been lenient in these cases even though any overall score
from an incomplete benchmarking suite is somewhat misleading.
All papers in our sample that use SPEC leave out the benchmarks written
in the Fortran language, instead using only the C and/or C++ ones.
We consider this to be justified because the prototypes built to test
the designs in these papers only support C and/or C++.
Moreover, it does not affect comparability
because this practice is widespread in the systems security
literature.
Another justified case of subsetting is the use of only C++
benchmarks for systems that do not apply to programs that are purely written in C.
Given that these systems would not be applied to C programs in practice,
their overhead on C has little meaning for their practicality.
In three cases, a subset of the benchmarks was omitted
because the system was based on a framework which
does not support them. We consider this acceptable if it is clearly indicated
because it is hard to avoid incompatibilities in third-party software.
Another case is the use of a subset
in a detailed evaluation after presenting overall numbers for the full set.
It is sensible to limit such an in-depth investigation
to the most interesting cases, generally those with most overhead,
and it provides more insight in which cases are hard for the system
to deal with without affecting comparability.
We have not marked any of the cases described in this paragraph
as a benchmarking crime because they are properly justified.

Although there can be legitimate reasons to select a subset of benchmarks,
we also found a large number of papers that did not properly justify
their subbenchmark selection. Four papers leave out a number of
SPEC subbenchmarks seemingly arbitrarily without even mentioning explicitly
that they were left out. This is a serious omission because these papers present
an overall overhead number that does not actually represent
the entire benchmarking suite, misleading readers into believing
that this number is directly comparable with those measured for other solutions.
While these subbenchmarks may have been left out for legitimate
reasons---for example they might not contain the type of memory safety bugs
that the system defends against---it is crucial to explain
why these particular benchmarks cannot be run with the system.
This not only justifies the lack of comparable numbers,
but also indicates the effectiveness or the limits of the solution
and helps competitors compare their solutions on these issues as well.

A second problem we found is leaving out the floating point
benchmarks of SPEC CPU without justification, which is a problem in four
of the papers in the sample. While this is not a random subset of SPEC CPU,
it is problematic because there are relatively many C++ benchmarks
in the floating point benchmarks. C++ programs tend to allocate
relatively many small heap objects, which stresses allocator instrumentation,
and contain relatively many virtual function calls,
which stresses indirect branch instrumentation. This means that for certain
classes of defenses, leaving out the floating-point programs is likely to result
in underestimating performance overhead.

Another problem we found in two papers that use SPEC is mixing subbenchmarks
from two different versions, namely SPEC CPU2000 and CPU2006.
While these benchmarking suites have some programs in common,
they use different workloads and their results are therefore not interchangeable.
The benchmarking suites are designed to be used as a balanced whole and
mixing versions results in unpredictable deviations in the overall results,
making those numbers incomparable.

One final problem that we found among the papers using SPEC CPU
is the use of an incorrect justification for leaving out subbenchmarks.
In particular, we found claims that some of the subbenchmarks
do not perform some instrumented operations while in reality they do.
Those subbenchmarks have thus been omitted in error, although the impact here is
less prominent than cases where benchmarks have been omitted arbitrarily since at
least the incompleteness of the benchmarking suite is clearly acknowledged.
Overall, we found a substantial number of cases where papers using SPEC CPU
improperly select a subset of the benchmarks and it seems plausible that
this has a substantial impact on the comparability of the results.

Not all papers use SPEC CPU to evaluate performance, although some do use
other standard benchmarking suites that test specific types of systems,
for example to evaluate the performance of operating
systems~\cite{unixbench,mcvoy1996lmbench} or browsers~\cite{octane,dromaeo}.
We found four such papers that use a subset of benchmarks
without justification. The impact in these cases is similar to those
where we found a subset of SPEC CPU is used.
In one additional case, a paper modified subbenchmarks without stating
why this was necessary. Like subsetting through selection,
this practice has a strong impact on comparability.

Papers that do not use a standard benchmarking suite generally use
a selection of supported programs and workloads for them
to measure performance. This is in itself acceptable as there
is not always a suitable benchmarking suite available.
A common example is the use of ApacheBench~\cite{apachebench}
to measure the performance of instrumented server programs.
However, even in these cases, it is important to justify selection
and avoid misrepresentation of the results.
We found five papers that presented a number of supported programs,
but then selected an unjustified subset of these programs
for benchmarking. This is problematic in cases where competing
solutions do include them, leaving the reader wondering which solution
would be faster, had the evaluation been more complete.

Another issue, which we found in one paper, is computing an overall overhead
figure over a number of self-selected programs. While this may be
useful to informally summarize overhead trends, it cannot be used as a reference
performance figure because such a figure strongly depends on the selection of the
programs. Instead, it would be more appropriate to provide a range of overheads
or always mention each program individually.

Finally, when defending against vulnerabilities,
it is important to ensure that the defense can prevent attacks in practice.
For this reason, many papers use vulnerabilities registered
in the CVE database~\cite{cve}. While this is an excellent way to assess
the effectiveness of defenses, authors generally select only a small number
of CVEs to evaluate their solution with. While this is understandable given the
often heroic effort, it is important to ensure that these CVE entries are
representative. We found five papers that lack a systematic selection
of vulnerabilities.
This means there is a risk of a biased selection,
masking limitations in the effectiveness of the solution being evaluated.

\subsubsection*{A3 - Selective data set hiding deficiencies}

Problems with selective data sets are not as common as
the other benchmarking crimes in this group,
with a total of 7 out of 48 applicable papers (15\%)
either committing the crime or being underspecified.
We found four papers where the impact of an important variable
is not considered in workload selection.
An example is not considering different levels of concurrency
when concurrency is expected to influence performance.
There are two papers in our sample where graphs suggest
that performance might reach a threshold but the range of the $x$-axis
is too limited to see it.

\subsection{Improper handling of benchmark results}

Improper handling of benchmark results is another very common group of
benchmarking crimes. 44 out of the 50 papers in our sample (88\%) commit
at least one of the five crimes in this group and two additional papers
do not provide enough information. However, it should be noted that this
is mostly due to lack of indication of significance (B4)
being very common in our sample.

\subsubsection*{B1 - Microbenchmarks representing overall performance}

Compared to the other benchmarking crimes, B1 stands out for being applicable
to relatively few papers because many papers do not present any microbenchmarks
at all. It is noteworthy that the use of microbenchmarks was much more common
in 2010 (14 out of 16 papers, 88\%) than in 2015 (10 out of 34 papers, 29\%).
Overall, this crime is committed in 6 out of 24 papers (25\%) and was more
common in 2010 even relative to the larger number of applicable cases.
In five cases, papers only present microbenchmarks and base their performance
claims on these microbenchmarks.
While there is one more paper that presents only microbenchmarks,
we have not labelled it as a crime since it only affects rare operations
that cannot realistically affect performance overhead on macrobenchmarks;
we consider it appropriate in cases where microbenchmarks can reveal overhead
that macrobenchmarks would not. Finally, one paper presents both microbenchmarks
and macrobenchmarks but bases its performance claims on the microbenchmarks
even though the macrobenchmarks show substantially more overhead.

\subsubsection*{B2 - Throughput degraded by x\% \(\Rightarrow\) overhead is x\%}

For most papers in our sample, this crime comes down to not ensuring that
the benchmark fully loads the CPU(s). This crime applies to
43 out of 50 papers (86\%), with the remainder not providing benchmarking
results that are intended to measure overhead. Out of these 43 papers,
16 commit the crime (37\%) and 2 are underspecified (5\%).
Most papers avoid this crime by either using a benchmarking suite known
to be CPU-bound or by ensuring that a manually constructed benchmark
fully loads the CPU, for example by running multiple concurrent threads
until all cores are fully loaded. Fifteen papers commit this crime by using
a benchmark that is not clearly CPU-bound without taking precautions to
ensure the CPU is fully loaded, while one other paper computes overhead
from latency rather than from throughput. In both cases, there is a substantial
risk that the actual overhead is underestimated because the overhead computation
does not consider the extra CPU load introduced by the protection mechanism
being evaluated.

\subsubsection*{B3 - Creative overhead accounting}

This crime is committed in 9 out of 50 papers (18\%) and it is impossible to
determine whether creative overhead accounting has been used
in 2 more papers (4\%). While this type of benchmarking crime is relatively
uncommon, it is very diverse. The most common variety is
to use magic numbers that are not supported by experiments
in overhead computations. In these cases, the results cannot be considered
methodologically sound. Another case of creative overhead accounting is
not considering some required instrumentations in the overhead numbers,
for example if the approach relies on the use of non-default compiler passes.
This results in an underestimation of the overhead that a user would experience
in practice. Another instance we found is to use percentage points
to compare overhead. For example, if solution A incurs 10\% overhead and
solution B incurs 20\% overhead then B has 100\% more overhead than A, not 10\%.
This misleads the reader into thinking that the differences are smaller
than they really are. One final issue we found is to mark overhead
as negligible because it is small compared to the standard deviation.
While this logic holds if the standard deviation is reasonable,
a large standard deviation is more likely to mean that the experiment
is set up incorrectly and the results are unreliable.
The proper reaction would be to improve the experiment to reduce measurement
error or, if this is not feasible, provide a confidence interval
on the overhead.
There is one more common issue with overhead computation,
namely computing an overall overhead when a number of subbenchmarks
have substantial negative overhead. We did not mark it as a crime
as it can be a legitimate effect of random measurement errors,
but we do want to raise the issue that it is important to explain
why overhead is negative for systems that should only decrease performance.
Large negative overheads can be an indication that the experiment is set up
incorrectly and authors should make an attempt to set up the experiment
in such a way as to reduce measurement errors.
Mytkowicz et al.~\cite{mytkowicz2009producing} provide guidelines
on how to achieve this. If negative overhead is simply ignored,
it may result in inaccurate performance numbers which are unsuitable
for comparison with competing solutions. In summary, creative overhead
accounting is a broad group of benchmarking crimes which can often result
in misleading and inaccurate results.

\subsubsection*{B4 - No indication of significance of data}

The lack of an indication of significance is a very widespread problem,
occurring in 38 out of 50 papers (76\%).
Two more papers contain qualitative significance statements without
putting a concrete upper bound on observed variations.
We expect papers that perform measurements that are subject
to random fluctuations, such as runtimes or throughput numbers,
to perform multiple runs to reduce standard errors and
to allow the standard deviation to be measured.
Papers should present the standard deviation or level of significance
for such numbers. We also accepted a general statement that ensures
that variation is at a very low level, such as
``all standard deviations are below 1\%''.

\subsubsection*{B5 - Incorrect averaging across benchmark scores}

We found that 12 out of 29 papers (41\%) incorrectly use the arithmetic mean
to average overhead numbers. For two more papers, there was
not enough information to decide whether the overall score was
computed correctly. This crime only applies to 29 out of the 50 papers (58\%)
because the remainder either does not present overhead numbers or
does not compute an overall score. One additional paper used
the arithmetic mean to average absolute overhead numbers,
which is acceptable and we did not count this as a crime.
Another paper presents overhead as a range,
which is also acceptable to obtain an overall indication of overhead.

\subsection{Using the wrong benchmarks}

Using the wrong benchmarks is a relatively rare group of benchmarking crimes,
with 14 out of the 50 papers in our sample (28\%) committing
at least one of the three crimes in this group and 5 additional papers
(10\%) not providing enough information. However, the crimes in this group
can have a major impact on the validity of the benchmarking results.

\subsubsection*{C1 - Benchmarking of simplified simulated system}

We found 5 out of 50 papers (10\%)
benchmarking a simplified simulated system. We did not count this as a crime
in cases where the use of a simplified system was explicitly acknowledged
and there was no practical way to avoid or compensate for it,
for example because the system relies on hardware that is not yet available.
In two of the papers that commit this crime we found that performance was
measured in a virtualized environment without need.
Virtualization does not incur a uniform slowdown,
but instead slows down operations that require an exit to the hypervisor
much more than unprivileged operations.
As a consequence, numbers measured in a virtual machine cannot be meaningfully
translated to numbers that would be measured on the bare metal.
Three other papers omitted some operations that would need to be performed
if the system were used in practice. Two of these cases were unjustified,
while the third had a good reason but did not consider
the impact on performance.

\subsubsection*{C2 - Inappropriate and misleading benchmarks}

The use of inappropriate and misleading benchmarks is moderately common,
with 8 out of 50 applicable papers (16\%) committing the crime and
2 more being underspecified. Although all the instances we found are
in papers published in 2015, the $\chi^2$-test reveals that this
can reasonably be the case due to mere chance ($p=0.198$).
Papers that make this mistake commonly also have a problem with
not evaluating potential performance degradation (benchmarking crime A1)
because inappropriate benchmarks often do not reveal important cases
where the system incurs overhead. The difference between the two is that
A1 applies if an important type of benchmark is missing even if
the included benchmarks are appropriate, while A3 can apply even if
some of the other benchmarks cover the relevant performance dimensions.

A typical example in the papers we surveyed includes the use of
IO-bound workloads in systems that introduce extra CPU load. This results in
benchmark results that suggest unrealistically low overhead. This poses a major
problem for later work, which is now expected to compare its performance against
the overly optimistic numbers measured before. Another situation is the case
where single-threaded single-process benchmarks are used to test systems where
concurrency is important, for example because they affect multiple cores. Like
in the previous case, the benchmark ignores an important part of the overhead.

One final problem is the use of performance benchmarks in cases
where high coverage is important, for example to detect false positives.
Since performance benchmarks are typically repetitive and do not test
error paths, they will not reach high coverage, revealing fewer false positives.

\subsubsection*{C3 - Same dataset for calibration and validation}

Unforuntately, this crime applies only to very few papers in our sample
(5 out of 50, 10\%). Out of these five, one papers commits the crime and
three are underspecified.

\subsection{Improper comparison of benchmarking results}

16 out of the 50 papers in our sample (32\%) commit at least one of the three
crimes that have to do with improper comparison of benchmarking results.
In addition, 6 more papers are underspecified with regards
to the criteria in this group.

\subsubsection*{D1 - No proper baseline}

This benchmarking crime stands out for having most papers by far
that are underspecified. There were 5 cases out of 50 (10\%)
where it was neither clear from the text of the paper nor implicit
from the context what baseline was used. This is an important problem,
not only because it means a reader cannot verify whether the baseline
is reasonable but also because it hampers reproducibility.
While the correct baseline should often be obvious,
it is always good to be explicit about it.

The benchmarking crime of not using a proper baseline was committed in
12 out of 50 papers (24\%). Five papers did not use a baseline at all,
presenting only raw numbers that do not give a good indication of overhead.
Another five papers used a nonstandard configuration for the baseline,
such as running on top of an instrumentation framework or using nonstandard
compiler options. In four of these papers, this likely means that performance
overhead is underestimated, while in a fifth the baseline was easier to attack
than a standard system would be.

One more paper used a simplification for the experimental system
without applying the same treatment to the baseline. A more appropriate approach
would be to measure both and use whichever approach is faster as the baseline.
Finally, we found a paper where a memory baseline is off by more than an order
of magnitude from the published reference baseline for the same benchmark.
This strongly suggests that it has been measured incorrectly and
such a difference requires an explanation in the paper.

\subsubsection*{D2 - Only evaluate against yourself}

The benchmarking crime of only evaluating against oneself applies
to relatively few papers, 15 out of 50 (30\%), because we only considered
those papers that actually perform a comparison. Out of those 15 papers,
2 (13\%) compare only against their own work while there would have been
appropriate alternatives.

\subsubsection*{D3 - Unfair benchmarking of competitors}

Like the previous benchmarking crime, this one only applies to
the 15 out of 50 papers (30\%) that actually perform a comparison.
The crime is committed in 4 out of 15 papers (27\%) and unclear due to
underspecification in 2 more papers. In two papers, we found competing solutions
were presented as having much higher overhead than in their original paper
with no explanation. Another paper selected an unoptimized number for
comparison, while an optimized version was also presented in the original paper.
In the fourth case, the configuration does not seem appropriate.

\subsection{Benchmarking omissions}

We found that 30 out of 50 papers (60\%) omit some important
benchmarking configurations, committing one or more of the four
benchmarking crimes in this group.

\subsubsection*{E1 - Not all contributions evaluated}

We found that 6 out of 50 papers do not evaluate all claimed contributions.
In particular, four papers do not test their effectiveness in securing programs
while two others do not evaluate the performance on some relevant applications.

\subsubsection*{E2 - Only measure runtime overhead}

While most papers evaluate runtime overhead, this is often not
the only relevant performance characteristic.
We found that 23 out of 50 papers (46\%) omitted other relevant
dimensions of performance. For almost all of these papers,
memory overhead is missing while it is reasonable to believe
the presented system might incur some memory overhead.
Given that memory usage can often be traded off against runtime performance
and that memory is a limited resource that must be shared between applications
running on a system, performance measurements are not complete
without measuring memory overhead. This means that, for example,
a paper that achieves similar runtime performance but greatly reduces
memory overhead compared to the state of the art is worth publishing.
If prior work lacks an evaluation of memory overhead,
it becomes harder to improve on it. Other missing measurements include
the amount of extra network and/or disk IO,
increased binary size after instrumentation,
and the time taken to instrument the protected program.
Like for memory, we have only counted cases where these performance dimensions
were not presented if there is a reasonable expectation that there
may be significant overhead.

\subsubsection*{E3 - False positives/negatives not tested}

For this benchmarking crime, we only considered papers where false positives
or negatives would be a potential problem. Out of the 19 papers where this
was the case, 6 papers (32\%) ignore the possibility of false positives or
negatives.

\subsubsection*{E4 - Elements of solution not tested incrementally}

This benchmarking crime only applies to the 25 out of 50 papers (50\%)
that actually consist of multiple components that could potentially
be measured independently or incrementally. From these papers,
4 (16\%) do not provide measurements for individual components.

\subsection{Missing information}

20 out of 50 papers (40\%) commit at least one of the four benchmarking crimes
in this group, leaving out some information that is important for completeness,
reproduction and/or sanity checking.

\subsubsection*{F1 - Missing platform specification}

11 out of 50 papers (22\%) do not provide a full specification of the hardware
used to run the benchmarks. Out of these, five do not give any information,
five more do not provide information about the networking setup and
the final one provides some information about the networking setup
but it is incomplete.

\subsubsection*{F2 - Missing software versions}

This benchmarking crime is committed by 12 out of 50 papers (24\%).
In six cases the paper does not specify the operating system used,
two papers do specify the operating system but not its version number,
one paper does not specify which hypervisor is used,
and three papers do not specify any information at all about the software
used to evaluate their systems.

\subsubsection*{F3 - Subbenchmarks not listed}

This benchmarking crime is applicable to the 38 out 50 papers (76\%)
which use subbenchmarks and, out of these, 7 (18\%) commit the crime.
While a seventh also does not list subbenchmark results explicitly,
the number of applications it was tested with is so large that presenting all of
them would be unpractical. Moreover, it does provide extensive statistics about
the subbenchmarks, which compensates for the missing information.
Therefore, we decided not to count it as having committed
this benchmarking crime. Still, it would have been even better if this paper
had discussed the methodology used to select the benchmarks that were used.
The other papers do not provide additional information
that can compensate for this lack of completeness.

\subsubsection*{F4 - Relative numbers only}

This benchmarking crime is applicable to 48 out 50 papers (96\%),
but none of these papers commit the crime in its worst form.
We found that 24 out of 48 applicable papers in our sample (50\%)
included only overheads.